\documentclass[12pt]{article}

\usepackage{scicite}

\usepackage{times}

\usepackage{graphicx}

\newenvironment{sciabstract}{%
\begin{quote} \bf}
{\end{quote}}

\newcommand{\mrm}{\mathrm}
\newcommand{\unit}{\,\mrm}
\newcommand{\degreeC}{\,^\circ\mathrm{C}}
\newcommand{\Th}{T_\mrm{h}}
\newcommand{\kT}{\kappa_\mrm{T}}
\newcommand{\kS}{\kappa_\mrm{S}}
\newcommand{\CP}{C_\mrm{P}}
\newcommand{\CV}{C_\mrm{V}}

\newcommand{\rhoLV}{\rho_\mrm{LV}}

\title{Compressibility anomalies in stretched water\\
and their interplay with density anomalies}

\author
{Vincent Holten,$^{1}$ Chen Qiu,$^{2}$ Emmanuel Guillerm,$^{1}$ Max Wilke,$^{3}$\\
Jaroslav Ri\v{c}ka,$^{2}$ Martin Frenz,$^{2}$ and Fr\'ed\'eric Caupin$^{1\ast}$\\
\\
\normalsize{$^{1}$Institut Lumi\`ere Mati\`ere, UMR5306 Universit\'e Claude Bernard Lyon
1-CNRS,}\\
\normalsize{Universit\'e de Lyon, 69622
Villeurbanne cedex, France}\\
\normalsize{$^{2}$Institute of Applied Physics, University of Bern, Sidlerstrasse 5, 3012
Bern, Switzerland}\\
\normalsize{$^{3}$Deutsches GeoForschungsZentrum and Universit\"{a}t Potsdam,}\\
\normalsize{Erd- und Umweltwissenschaften, Karl-Liebknecht-Str. 24-25, 14473 Potsdam, Germany}\\
\\
\normalsize{$^\ast$To whom correspondence should be addressed; E-mail:  frederic.caupin@univ-lyon1.fr.}
}

\date{}

\begin{document} 




\maketitle

%

\begin{sciabstract}
Water keeps puzzling scientists because of its numerous properties which behave oppositely to usual liquids: for instance, water expands upon cooling, and liquid water is denser than ice. To explain this anomalous behaviour, several theories have been proposed, with different predictions for the properties of supercooled water (liquid at conditions where ice is more stable). However, discriminating between those theories with experiments has remained elusive because of spontaneous ice nucleation. Here we measure the sound velocity in liquid water stretched to negative pressure, and derive an experimental equation of state, which reveals compressibility anomalies. We show by rigorous thermodynamic relations how these anomalies are intricately linked with the density anomaly. Some features we observe are necessary conditions for the validity of two theories of water.
\end{sciabstract}


Liquid water exhibits numerous anomalies and different scenarios have been proposed to explain them\cite{Gallo_water_2016}. In particular, the existence of a line of density maxima along isobars has been related to putative maxima in compressibility\cite{Sastry_singularityfree_1996} and heat capacity\cite{Poole_density_2005}. These maxima may arise from an intriguing phase separation of water in two distinct liquids~\cite{Poole_phase_1992,Gallo_water_2016}, although they can also be explained without resorting to such a phase separation~\cite{Sastry_singularityfree_1996}. However, the compressibility and heat capacity maxima, whose existence is predicted by molecular dynamics simulations\cite{Poole_density_2005,Gonzalez_comprehensive_2016}, have hitherto not been observed in experiments. Alternative theoretical scenarios, namely the stability limit conjecture~\cite{Speedy_stabilitylimit_1982} and the critical-point-free scenario~\cite{Poole_effect_1994,Angell_insights_2008}, do not require the existence of compressibility and heat capacity maxima, but rather predict a divergence of these quantities at low temperature.

Another type of anomaly, namely a minimum in sound velocity along one isochore, was recently discovered at negative pressure~\cite{Pallares_anomalies_2014,Pallares_equation_2016}. At negative pressure, the liquid is mechanically stretched, in a state metastable with respect to vapor. To date, the only method able to reach significantly negative pressures (beyond $-100\unit{MPa}$) uses $3-10\unit{\mu m}$ fluid inclusions (FIs) of water in a quartz crystal, and stretching is obtained by cooling liquid water at nearly constant volume~\cite{Zheng_liquids_1991,Azouzi_coherent_2012}.
In our previous work~\cite{Pallares_anomalies_2014,Pallares_equation_2016}, we could only study two FIs along different isochores, and only one clearly showed a minimum in sound velocity. In the present work, we have measured more FIs showing a sound velocity minimum, and reached more negative pressures. We have thus established a more accurate experimental equation of state (EoS) down to $-137\unit{MPa}$. The new EoS strongly supports the existence of the compressibility maxima predicted by some theories of water. Furthermore, we establish new thermodynamic relations between the line of density maxima and the sound velocity anomalies. In contrast to previous works, this provides a relation between quantities that are directly observable in experiments. The corresponding lines of extrema obtained from our experimental data run on a "collision course"\cite{Debenedetti_physics_2013}, which suggests that the line of density maxima reaches a maximum temperature around $-150\unit{MPa}$, a feature compatible with only two of the proposed scenarios for water.

\begin{figure}
\centerline{\includegraphics[width=15cm]{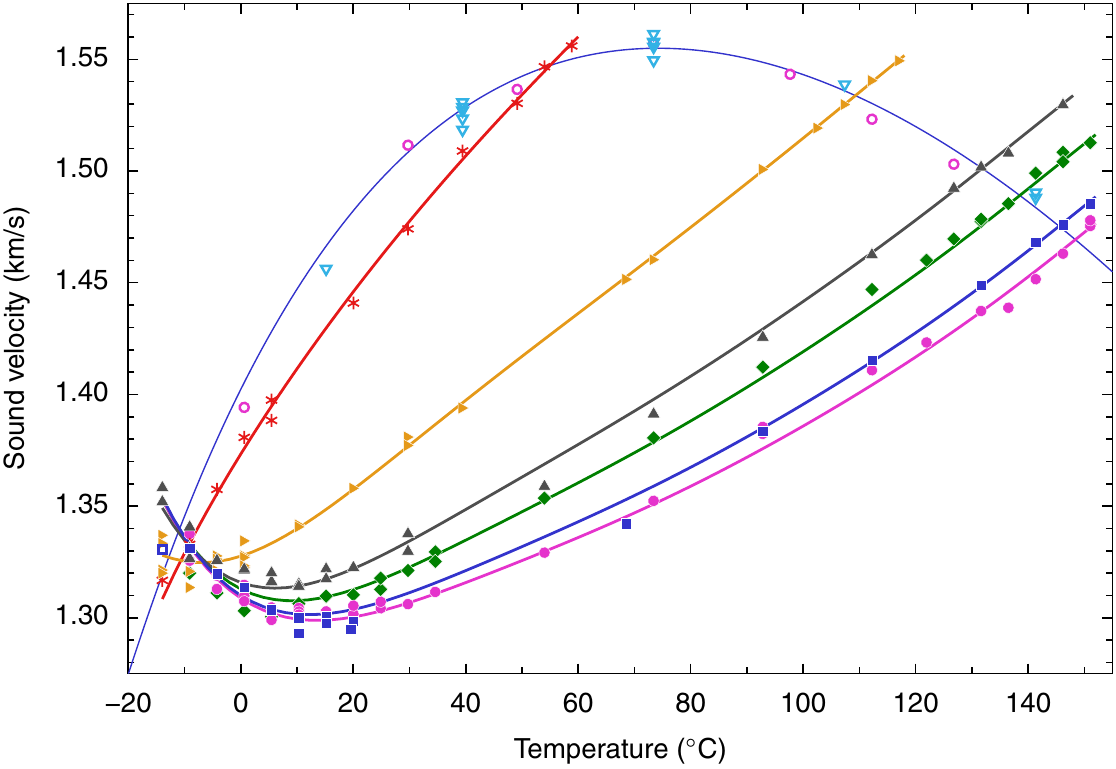}}
\caption{Sound velocity as a function of temperature in several fluid inclusions of water in quartz. The open symbols show data taken when a bubble is present, which follow the thin blue solid curve known for water at the liquid-vapor equilibrium~\cite{Wagner_IAPWS_2002}. The filled symbols show data taken for inclusions full of liquid. The thick solid curves through the filled symbols show the interpolation of $c(T,\rho)$ used to construct the EoS\cite{Pallares_equation_2016} (see Supplementary Materials for details).}
\label{fig:c-T}
\end{figure}

Six FIs synthesized hydrothermally~\cite{Qiu_exploration_2016} were selected to cover evenly the density range between liquid-vapor equilibrium down to the cavitation limit, at which vapor nucleation occurs spontaneously. Sound velocity was measured from $-13.9$ to $151\degreeC$ using the Brillouin micro-spectrometer previously described~\cite{Pallares_anomalies_2014}. The data points are shown in Fig.~\ref{fig:c-T}. When the FI contains a bubble, the sound velocity measured along liquid-vapor coexistence coincides with the known value~\cite{Wagner_IAPWS_2002}. Upon heating, the vapor bubble disappears at the homogenization temperature $\Th$ and the FI contains only liquid, at a density calculated from the observed $\Th$ (a small density correction is made at other temperatures to account for the compliance of the quartz matrix~\cite{Pallares_equation_2016}). Upon cooling, water remains liquid but occupies a volume which is larger than for stable liquid water, thus becoming metastable with respect to the vapor. The sound velocity is then lower than at liquid-vapor coexistence, except at the lowest temperatures as previously observed~\cite{Pallares_anomalies_2014}. The five FIs with the highest $\Th$ values (lowest densities) clearly show a minimum in the sound velocity vs. temperature. The temperature of this minimum increases with increasing $\Th$, i.e. with decreasing density. 

\begin{figure}
\centerline{\includegraphics[width=15cm]{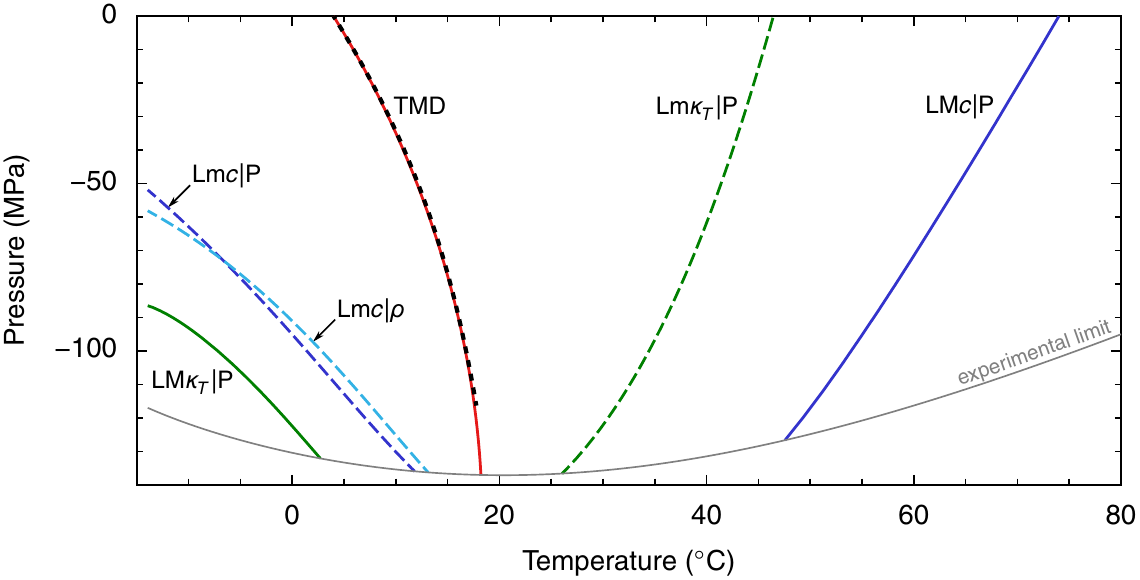}}\caption{Lines of extrema in thermodynamic quantities obtained from the sound velocity data, plotted down to the limit reached in our experiment. The line of minimum sound velocity along isochores (Lm$c|\rho$) reflects the minima seen in the curves of Fig.~\ref{fig:c-T}. The TMD (solid red line) confirms and extends our previous determination (short dashed black line)~\cite{Pallares_anomalies_2014}. The present data provide strong evidence in favor of the existence of a line of maximum compressibility along isobars (LM$\kT|P$). The lines are in the order required by thermodynamics, from left to right: LM$\kT|P$, Lm$c|P$, TMD line, Lm$\kT|P$, and finally LM$c|P$. Lines of $\kS$ extrema along isobars are not shown for clarity, but also obey the correct thermodynamic order.}
\label{fig:lines}
\end{figure}

In order to generate an experimental EoS from our data, we proceed as in Ref.~\cite{Pallares_equation_2016}. We first interpolate the sound velocity data (solid curves in Fig.~\ref{fig:c-T}), and then integrate the appropriate thermodynamic relations to get pressure $P$ and isothermal compressibility $\kappa_T$ (see Materials and Methods). The calculated EoS reveals a series of thermodynamic anomalies, which are displayed in Fig.~\ref{fig:lines}. The line of sound velocity minima along isochores (Lm$c|\rho$) corresponds to the $c(T)$ minima shown in Fig.~\ref{fig:c-T}. The EoS allows plotting other lines of sound velocity anomalies along isobars: maxima (LM$c|P$) prolonging to negative pressure the line known to exist at positive pressure, but also a new line of sound velocity minima (Lm$c|P$). The temperature of maxima of density along isobars (TMD), which occurs at $4\degreeC$ at ambient pressure, monotonically increases to $18\degreeC$ when the pressure is decreased to $-137\unit{MPa}$. This confirms and extends our previous results~\cite{Pallares_equation_2016} which were limited to $-116\unit{MPa}$. Unfortunately, while the TMD gets more and more vertical as the pressure is lowered, it does not reach a turning point in the measured range. This would be decisive in discriminating between the scenarios proposed to explain water anomalies, as two of them predict a turning point\cite{Poole_phase_1992,Sastry_singularityfree_1996}, whereas two others predict a monotonic TMD line\cite{Speedy_stabilitylimit_1982,Poole_effect_1994,Angell_insights_2008}. One would need to obtain data at more negative pressure, but this is not possible with our technique: our lowest density FI has reached the homogeneous cavitation line measured in Ref.~\cite{Qiu_exploration_2016}.

Our work does however give an answer to the question about the behaviour of isothermal compressibility $\kappa_T$: we find a line of maxima of $\kT$ along isobars (LM$\kT |P$) (Fig.~\ref{fig:lines}). The existence of such maxima was already considered in our previous work~\cite{Pallares_equation_2016}, but no clear conclusion could be reached at that time, because small changes to the interpolation function made the maximum appear or disappear. Here, thanks to the larger number of samples and the lowest pressure reached, the existence of a LM$\kT |P$ is unambiguously established, being now robust against changes in the choice of interpolation and against arbitrary reduction of our data set (see Supplementary Materials for details). The existence of a LM$\kT |P$ is key to the debate about the origin of the anomalous behaviour of water. Indeed, the second critical point~\cite{Poole_phase_1992} and singularity-free~\cite{Sastry_singularityfree_1996} scenarios necessarily require the existence of a LM$\kT |P$, which would extend up to positive pressure and explain the increase in $\kT$ upon cooling water into the supercooled region at positive pressure. In contrast, the stability limit conjecture~\cite{Speedy_stabilitylimit_1982} and the critical-point free scenario~\cite{Angell_insights_2008} explain this increase in $\kT$ by a divergence in $\kT$ at lower temperature, and do not require a LM$\kT |P$. Note however that, even for these two latter scenarios, the possibility of a LM$\kT |P$ ending by merging with a second Lm$\kT |P$ at lower temperature than our experimental range, although not required, cannot be excluded. To further discuss the possible shape of the TMD line, we now focus on the relation between density and sound velocity anomalies.



Sound velocity $c$ is related to the adiabatic compressibility $\kS$ through the Newton-Laplace relation: $c=1/\sqrt{\rho \kS}$, where $\rho$ is the density. Anomalies in $c$ are thus related to anomalies in $\kS$. Here we make the connection between anomalies in $c$ and $\kS$ and the TMD, with an argument similar to that used for $\kT$ in Ref.~\cite{Sastry_singularityfree_1996}, where it was proven that:
\begin{equation}
\left( \frac{\partial \kT}{\partial T} \right)_P = \frac{1}{V} \left(\frac{\partial^2 V}{\partial T^2}\right) \left(\frac{\mrm{d}P}{\mrm{d}T} \right)_\mrm{TMD}^{-1} \quad \textrm{along the TMD line}\; .
\label{eq:Sastry}
\end{equation}
Here $V$ is the molar volume, and $(\mrm{d}P/\mrm{d}T)_\mrm{TMD}$ is the slope of the TMD line in the $T-P$ plane. The details of our derivation are given in the Supplementary Materials, and lead to the following relations valid along the TMD line:
\begin{equation}
\left( \frac{\partial \kS}{\partial T} \right)_P = \left( \frac{\partial \kS}{\partial T} \right)_V = \frac{1}{\rho} \left( \frac{\partial (1/c^2)}{\partial T} \right)_V = \frac{1}{\rho} \left( \frac{\partial (1/c^2)}{\partial T} \right)_P = \frac{1}{V} \left(\frac{\partial^2 V}{\partial T^2}\right) \left(\frac{\mrm{d}P}{\mrm{d}T} \right)_\mrm{TMD}^{-1} \; ,
\label{eq:mastereq}
\end{equation}
At the TMD line, $V$ reaches a minimum vs. $T$ at constant $P$, therefore $(\partial^2 V/\partial T^2)_P>0$. Equation~\ref{eq:mastereq} and the presence of a TMD with negative slope $(\mrm{d}P/\mrm{d}T)_\mrm{TMD}<0$ immediately show that $\kS$ must decrease and $c$ increase when the temperature is increased at constant pressure or volume starting from the TMD. This is a behaviour opposite to that of normal liquids, in which $\kS$ increases and $c$ decreases when heating along isobars. In water, the negatively sloped TMD thus causes the anomalous behaviour. At pressures below the liquid-vapor critical pressure, increasing temperature eventually brings the liquid on the spinodal line where it becomes unstable. It is known~\cite{Ahlers_chap_1976,Speedy_stabilitylimit_1982} that on the spinodal line, $\kT$, the isobaric heat capacity $\CP$, and the isobaric expansion coefficient $\alpha_P$ diverge, whereas $\kS$, the isochoric heat capacity $\CV$ and the thermal pressure coefficient $\beta=(\partial P/\partial T)_V$ remain finite. The divergence of $\kT$ on the spinodal, together with Eq.~\ref{eq:Sastry}, was used in Ref.~\cite{Sastry_singularityfree_1996} to conclude about the existence of a line of minima of $\kT$ along isobars (Lm$\kT|P$) in water. Because $\kS$ does not diverge on the spinodal, we cannot use the same reasoning. However, a detailed consideration of the behaviour of thermodynamic quantities near the liquid-vapor critical point allows us to rigorously prove that the negatively sloped TMD line implies the existence of lines of extrema of $\kS$ and $c$ along isobars. Furthermore, for the case of water, we specify the nature of these lines, showing the existence of a line of minima in $\kS$ along isobars (Lm$\kS|P$) and a line of maxima of $c$ along isobars (LM$c|P$) (see Supplementary Materials for the proof). Both lines are to the right of the TMD line in the $T-P$ plane. A LM$c|P$ is indeed well known for water at positive pressure. Our data extends the LM$c|P$, finding that the maximum in $c$ occurs at lower temperatures when pressure is decreased (see Fig.~\ref{fig:lines}).

Figure~\ref{fig:lines} also displays other lines of extrema revealed by our experiments: sound velocity minima along isochores (Lm$c|\rho$) and isobars (Lm$c|P$). All the extrema lines we observe approach the TMD line when the pressure decreases. It is known that, if the TMD intersects the line of $\kT$ extrema along isobars, then the TMD must reach at the intersection a maximum temperature (see Eq.~\ref{eq:Sastry} and Ref.~\cite{Sastry_singularityfree_1996}). The same holds for the intersection with lines of $\kS$ and $c$ extrema along isobars and along isochores. Indeed, Eq.~\ref{eq:mastereq} shows that at the intersection, as the $T$ derivatives of $\kS$ and $1/c^2$ vanish, the TMD slope $(\mrm{d}P /\mrm{d}T )_\mrm{TMD}$ becomes infinite. In the present work we find a bundle of lines converging towards the TMD. They all run on a ``collision course''~\cite{Debenedetti_physics_2013}. Interestingly, the order of these lines is given by thermodynamics (see Supplementary Material): with increasing temperature along an isobar, we first meet LM$\kT|P$, then Lm$c|P$, TMD line, Lm$\kT|P$, and finally LM$c|P$. This shows that Lm$c|P$ is more accessible to experiments than LM$\kT|P$, and that the closest bounds to the predicted TMD turning point from extrema along isobars are given by Lm$c|P$ and Lm$\kT|P$. Our experiment confirms this, and now brackets the possible location of the TMD turning point in a $15\unit{K}$ interval. This interval can be further reduced to $13\unit{K}$ by noticing that, for our EoS, Lm$c|\rho$ is even closer to the TMD than the Lm$c|P$ (Fig.~\ref{fig:lines}).

\begin{figure}
\centerline{\includegraphics[width=15cm]{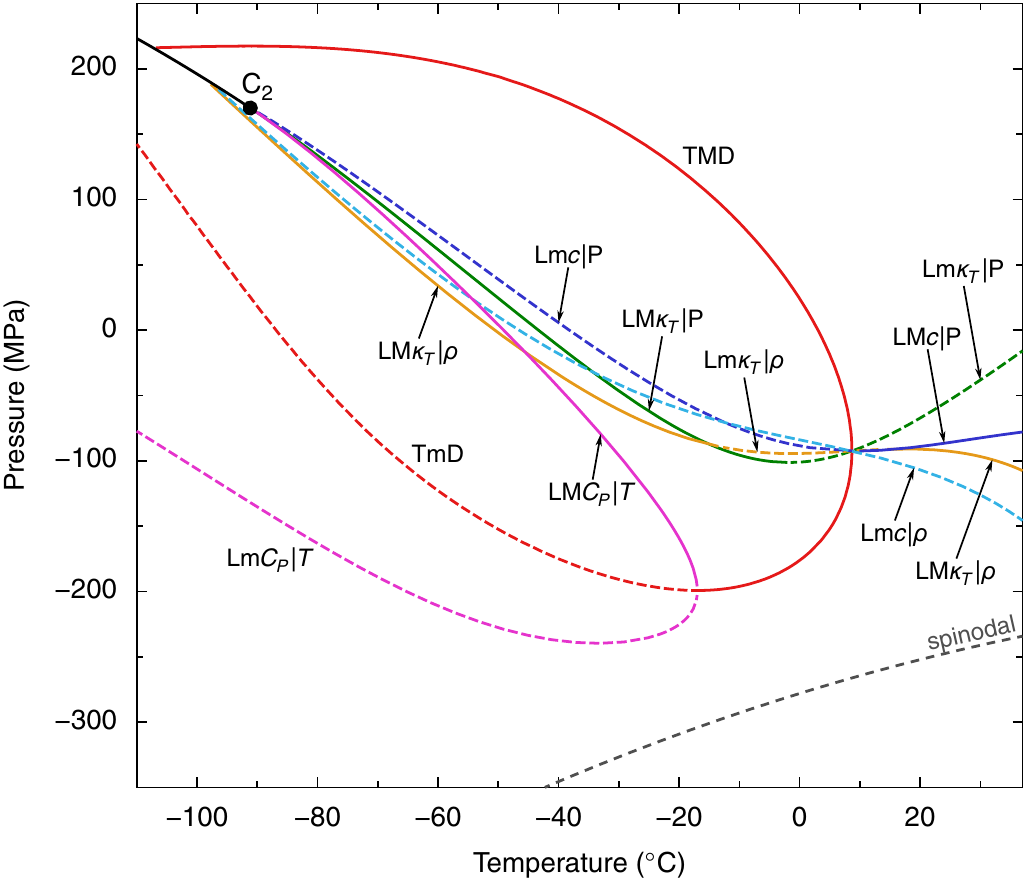}}\caption{Lines of extrema in thermodynamic quantities obtained from the two-structure EoS of the TIP4P/2005 potential for water~\cite{Biddle_twostructure_2017}. The liquid-liquid transition (solid black line) ending at a liquid-liquid critical point (C$_2$), and the spinodal line (dashed gray line) for this model are also shown. In addition to the LM$\kT|P$, TMD line, Lm$\kT|P$, LM$\CP|T$ and Lm$\CP|T$ already reported, we have calculated the location of the Lm$c|P$, LM$c|P$, Lm$c|\rho$, Lm$\kT|\rho$, and LM$\kT|\rho$ lines. Lines of $\kS$ extrema along isobars are also found, but not shown for clarity.}
\label{fig:TIP4P2005}
\end{figure}

Molecular dynamics simulations with a realistic potential for water provide a means to access a larger degree of metastability than experiments. They can be used to locate the lines of extrema in thermodynamic functions, in order to compare them with our experimental and theoretical findings. Recently, simulations of the TIP4P/2005 potential for water have been parameterized down to the liquid-vapor spinodal with a two-state model that provides analytic formulas for all thermodynamic properties~\cite{Biddle_twostructure_2017}. The model reproduces well the lines of extrema that were directly simulated for TIP4P/2005, and allow to compute other lines such as LM$c|P$ or Lm$c|P$. The results are shown in Fig.~\ref{fig:TIP4P2005}. The model system based on the TIP4P/2005 potential captures all the lines of extrema obtained from our experiments in the correct order. In addition, the lines can be followed closer to the spinodal. This allows observing the TMD turning point, and the intersection with other lines there, as predicted by Eq.~\ref{eq:mastereq}. New lines of extrema are also found: maxima and minima in $\kT$ along isochores (LM$\kT|\rho$ and Lm$\kT|\rho$, respectively). Note that two points on the LM$\kT|\rho$ (not shown) were previously found in simulations~\cite{Pallares_anomalies_2014}, and lie close to our calculated line. Using more thermodynamic relations, we show in the Supplementary Material that the line of extrema in $\kT$ is caused by the positively sloped TMD (below its turning point), and intersects the TMD at its turning point.

As anticipated before~\cite{Pallares_anomalies_2014,Caupin_escaping_2015}, the study of water in the doubly metastable region, both supercooled and at negative pressure, allows accessing the LM$\kT|P$ which has been sought for decades, and whose existence we have now established. At ambient pressure, a maximum in $\kT$ was never found above $-38\degreeC$, the usual experimental limit of supercooling with standard techniques, but might become accessible with the recently used technique of ultrafast cooling of water droplets in vacuum~\cite{Sellberg_ultrafast_2014}. Indeed, observation of liquid water at a calculated temperature as low as $-46\degreeC$ has been reported~\cite{Sellberg_ultrafast_2014}. Although more recent experiments~\cite{Goy_shrinking_2017} indicate that this temperature might have been underestimated and might be in fact above $-42.6\degreeC$, further investigations are needed, for instance to obtain $\kT$ from the limiting value at zero wavevector of small-angle x-ray or neutron scattering~\cite{Clark_smallangle_2010}. Furthermore, the relations between different lines of extrema we have proven, being based on general thermodynamic relations, are applicable to any liquid with a TMD line. The properties we have found, together with previous results~\cite{Sastry_singularityfree_1996,Poole_density_2005}, thus constitute a general framework to describe and understand the anomalies of other liquids presenting a density maximum.

\bibliography{articles}
\bibliographystyle{Science}

\section*{Acknowledgments}
We acknowledge the use of the HP-GeoMatS Lab of the Geo.X partners GFZ and Universit\"{a}t Potsdam for pure water fluid inclusion synthesis. This work was supported by Agence Nationale de la Recherche through project CGS$\mu$Lab-ANR-12-SEED-0001, and by the Swiss National Science Foundation [grant number 200021-140777].

\newpage
\section*{Supplementary materials}

\hspace{6mm}\textbf{Materials and Methods}

Fluid inclusions were produced by hydrothermal synthesis\cite{Bodnar_synthetic_1985}, selecting the temperature and pressure conditions to reach densities covering the required range. Brillouin spectroscopy was performed and analyzed as described elsewhere~\cite{Pallares_anomalies_2014}. The equation of state was obtained by thermodynamic integration following the method detailed in Ref.~\cite{Pallares_equation_2016}.\\

\textbf{Supplementary Text}

\textbf{Sensitivity of the results on the choice of interpolation and data set}

Similar to Ref.~\cite{Pallares_equation_2016}, we have tested a number of interpolations of the $c(T,\rho)$ data such as the one shown in Fig.~\ref{fig:c-T}. To further test how the results depend on the choice of a particular interpolation, we have also introduced a different class of interpolation. Namely, we have expanded $c$ as powers of $\rho-\rhoLV(T)$, with $\rhoLV(T)$ the density of the liquid in equilibrium with its vapor, in two manners:
\begin{eqnarray}
c(T,\rho) &=& c(T,\rhoLV(T)) + \left(\frac{\partial c}{\partial \rho} \right)_T \!(T,\rhoLV(T)) [ \rho-\rhoLV(T) ]\nonumber \\
& & + a_2 (T) [ \rho-\rhoLV(T) ]^2 + a_3 (T) [ \rho-\rhoLV(T) ]^3 , \label{eq:interprho1}\\ 
c(T,\rho) &=& c(T,\rhoLV(T)) + \left(\frac{\partial c}{\partial \rho} \right)_T \!(T,\rhoLV(T)) [ \rho-\rhoLV(T) ] + a(T) [ \rho-\rhoLV(T) ]^{b(T)} .\label{eq:interprho2}
\end{eqnarray}
Here the constants appearing in the two first terms of the right hand side of each equation are known from the accurate data for stable water above $\rhoLV(T)$~\cite{Wagner_IAPWS_2002}. Several functional forms were tested for the free parameters $a_2(T)$ and $a_3(T)$ on the one hand (Eq.~\ref{eq:interprho1}), and $a(T)$ and $b(T)$ on the other hand (Eq.~\ref{eq:interprho2}).

We have selected 2 interpolations with Eq.~\ref{eq:interprho1}, and 2 interpolations with Eq.~\ref{eq:interprho2}, with slight qualitative and quantitative differences in their description of the data. They all show a good quality of fit (reduced $\chi^2<$0.8). Similar to Ref.~\cite{Pallares_equation_2016}, we have compared the lines of extrema in thermodynamic properties for the four choices of interpolation. The only notable difference is that, whereas for the 2 interpolations with Eq.~\ref{eq:interprho1} LM$c|P$ is monotonic as in Fig.~\ref{fig:lines}, for the 2 interpolations with Eq.~\ref{eq:interprho2} LM$c|P$ reaches a minimum temperature, and turns back to higher temperatures for lower pressures. This is allowed by thermodynamics for all the theoretical scenarios we are discussing. For all other lines, the results are only slightly dependent on the choice of interpolation. In particular, the existence and location of the LM$\kT|P$ is a robust feature.

Using the best interpolation with Eq.~\ref{eq:interprho1} used to plot Figs.~\ref{fig:c-T} and \ref{fig:lines}, we have performed another sensitivity test. We have repeated the whole procedure on two reduced data sets, each generated by removing data for one sample (the second and third less dense sample, respectively). All lines of thermodynamic extrema remain nearly the same as for the whole data set.
\\

\textbf{Derivation of Eq.~\ref{eq:mastereq}}

This section provides the proof of the 4 equalities appearing in Eq.~\ref{eq:mastereq}. On the TMD line, by definition,
\begin{equation}
\left( \frac{\partial V}{\partial T} \right)_P = 0 .
\label{eq:dVdT_P1}
\end{equation}
Therefore, at the TMD,
\begin{equation}
\left( \frac{\partial V}{\partial S} \right)_P= \left( \frac{\partial V}{\partial T} \right)_P \left( \frac{\partial T}{\partial S} \right)_P = \left( \frac{\partial V}{\partial T} \right)_P \frac{T}{\CP} = 0\; .
\label{eq:dVdS_P}
\end{equation}
Using again Eq.~\ref{eq:dVdT_P1}, the second derivative $(\partial^2 V/\partial S^2)_P$ can be simplified at the TMD:
\begin{equation}
\left( \frac{\partial^2 V}{\partial S^2} \right)_P= \left( \frac{\partial^2 V}{\partial T^2} \right)_P \left(\frac{T}{\CP}\right)^2 + \left( \frac{\partial V}{\partial T} \right)_P \left( \frac{\partial (T/\CP )}{\partial S} \right)_P = \left( \frac{\partial^2 V}{\partial T^2} \right)_P \left(\frac{T}{\CP}\right)^2 \; .
\label{eq:d2VdS2_P}
\end{equation}

Differentiating Eq.~\ref{eq:dVdS_P} valid along the TMD line with respect to the variables $S$ and $P$, we find:
\begin{equation}
\mrm{d} \left[ \left(\frac{\partial V}{\partial S} \right)_P \right] = 0 = \left(\frac{\partial^2 V}{\partial S^2} \right)_P \mrm{d}S + \frac{\partial^2 V}{\partial P \partial S} \mrm{d}P \; ,
\label{eq:dVdS_P2}
\end{equation}
so that in the $S-P$ plane, the TMD slope is:
\begin{equation}
\left( \frac{\mrm{d}P}{\mrm{d}S} \right)_\mrm{TMD} = -
\left(\frac{\partial^2 V}{\partial S^2}\right)_P
\left(\frac{\partial^2 V}{\partial P \partial S} \right)^{-1} \; .
\label{eq:dPdS_TMD}
\end{equation}
It is more useful to write the slope in the $T-P$ plane. For this we write:
\begin{equation}
\mrm{d}T = \left( \frac{\partial T}{\partial S} \right)_P \mrm{d}S + \left( \frac{\partial T}{\partial P} \right)_S \mrm{d}P \; .
\label{eq:dT_S-P}
\end{equation}
Along the TMD line, Eq.~\ref{eq:dT_S-P} becomes:
\begin{equation}
\left( \frac{\mrm{d} T}{\mrm{d} P} \right)_\mrm{TMD} =
\frac{T}{\CP} \left( \frac{\mrm{d} S}{\mrm{d} P} \right)_\mrm{TMD} + \left( \frac{\partial T}{\partial P} \right)_S \, .
\label{eq:dTdP_TMD}
\end{equation}
Using standard thermodynamic relations, we write:
\begin{equation}
\left( \frac{\partial T}{\partial P} \right)_S = \frac{T V \alpha_P}{\CP} = 0 \;\textrm{at the TMD} \; .
\label{eq:dTdP_S}
\end{equation}
Therefore, Eqs.~\ref{eq:dPdS_TMD}, \ref{eq:dTdP_TMD} and \ref{eq:dTdP_S} yield:
\begin{equation}
\left( \frac{\mrm{d} P}{\mrm{d} T} \right)_\mrm{TMD} = \frac{\CP}{T} \left( \frac{\mrm{d} P}{\mrm{d} S} \right)_\mrm{TMD} = - \frac{\CP}{T} \left(\frac{\partial^2 V}{\partial S^2}\right)_P
\left(\frac{\partial^2 V}{\partial P \partial S} \right)^{-1} \; .
\label{eq:dPdT_TMD}
\end{equation}

From the definition of the adiabatic compressibility:
\begin{equation}
\kS = - \frac{1}{V} \left( \frac{\partial V}{\partial P} \right)_S \; ,
\label{eq:kS}
\end{equation}
we get:
\begin{equation}
\left( \frac{\partial \kS}{\partial S} \right)_P = \frac{1}{V^2} \left(\frac{\partial V}{\partial S} \right)_P \left(\frac{\partial V}{\partial P} \right)_S - \frac{1}{V} \frac{\partial^2 V}{\partial S \partial P} \; .
\label{eq:dkSdS_P}
\end{equation}
At the TMD, Eq.~\ref{eq:dVdS_P} is valid, and Eq.~\ref{eq:dkSdS_P} simplifies to:
\begin{equation}
\left( \frac{\partial \kS}{\partial S} \right)_P = - \frac{1}{V} \frac{\partial^2 V}{\partial P \partial S} \; .
\label{eq:dkSdS_P2}
\end{equation}
Combining Eq.~\ref{eq:dPdT_TMD} and \ref{eq:dkSdS_P2}, we find:
\begin{equation}
\left( \frac{\partial \kS}{\partial S} \right)_P = \frac{\CP}{T V} \left( \frac{\partial^2 V}{\partial S^2} \right)_P \left(\frac{\mrm{d}P}{\mrm{d}T} \right)_\mrm{TMD}^{-1} \; .
\label{eq:dkSdS_P3}
\end{equation}
At the TMD, using Eqs.~\ref{eq:d2VdS2_P} and \ref{eq:dkSdS_P3}, we thus obtain one of the relations written in Eq.~\ref{eq:mastereq}:
\begin{equation}
\left( \frac{\partial \kS}{\partial T} \right)_P = \left( \frac{\partial S}{\partial T} \right)_P \left( \frac{\partial \kS}{\partial S} \right)_P = \frac{1}{V} \left( \frac{\CP}{T} \right)^2 \left( \frac{\partial^2 V}{\partial S^2} \right)_P \left(\frac{\mrm{d}P}{\mrm{d}T} \right)_\mrm{TMD}^{-1} = \frac{1}{V} \left( \frac{\partial^2 V}{\partial T^2} \right)_P \left(\frac{\mrm{d}P}{\mrm{d}T} \right)_\mrm{TMD}^{-1} \; .
\label{eq:dkSdT_P}
\end{equation}
For the derivative at constant volume, we use:
\begin{equation}
\left( \frac{\partial \kS}{\partial T} \right)_V = \left( \frac{\partial \kS}{\partial T} \right)_P + \left( \frac{\partial \kS}{\partial P} \right)_T \left(\frac{\partial P}{\partial T} \right)_V \; .
\label{eq:dkSdT_V}
\end{equation}
From Eq.~\ref{eq:dVdT_P1},
\begin{equation}
\left( \frac{\partial P}{\partial T} \right)_V = - \left( \frac{\partial P}{\partial V} \right)_T \left( \frac{\partial V}{\partial T} \right)_P = 0 \;\textrm{at the TMD} \; ;
\label{eq:dPdT_V}
\end{equation}
an isochore reaches its lowest pressure at the TMD. At the TMD Eq.~\ref{eq:dkSdT_V} thus becomes:
\begin{equation}
\left( \frac{\partial \kS}{\partial T} \right)_V = \left( \frac{\partial \kS}{\partial T} \right)_P \; ,
\label{eq:dkSdT_V2}
\end{equation}
which gives another relation written in Eq.~\ref{eq:mastereq}.

From the Newton-Laplace relation: $c=1/\sqrt{\rho \kS}$, we see that along an isochore for which $V$ and $\rho$ are constant:
\begin{equation}
\left( \frac{\partial (1/c^2)}{\partial T} \right)_V = \rho \left( \frac{\partial \kS}{\partial T} \right)_V \, .
\label{eq:dinvc2dT_V}
\end{equation}
Finally, to study the temperature derivative of $c$ along an isobar, we write:
\begin{equation}
\left( \frac{\partial (1/c^2)}{\partial T}\right)_P = \left( \frac{\partial \rho}{\partial T}\right)_P \kS + \rho \left( \frac{\partial \kS}{\partial T}\right)_P \; ,
\label{eq:dinvc2dT_P}
\end{equation}
which at the TMD where $(\partial \rho / \partial T)_P = 0$ reduces to:
\begin{equation}
\left( \frac{\partial (1/c^2)}{\partial T}\right)_P = \rho \left( \frac{\partial \kS}{\partial T}\right)_P \; ,
\label{eq:dinvc2dT_P2}
\end{equation}
which completes the proof of Eq.~\ref{eq:mastereq}.\\

\textbf{Existence of lines of extrema in $\kS$ and $c$ along isobars above the TMD (Lm$\kS|P$ and LM$c|P$ in the case of water)}

Here we consider a liquid which exhibits a negatively-sloped TMD line in the $T-P$ plane. Reference~\cite{Sastry_singularityfree_1996} used Eq.~\ref{eq:Sastry} and the divergence of $\kT$ on the liquid-to-vapor spinodal to conclude about the existence of a line of minima of $\kT$ along isobars (Lm$\kT|P$) in that case. Because $\kS$ does not diverge and $c$ does not vanish on the spinodal, we cannot use the same reasoning.

Eq.~\ref{eq:mastereq} shows that, along the negatively sloped TMD line:
\begin{equation}
\left( \frac{\partial \kS}{\partial T} \right)_P < 0 \quad\mrm{and}\quad \left( \frac{\partial c}{\partial T} \right)_P >0 \; .
\label{eq:dkSdT-dcdT}
\end{equation}
Let us consider how these derivatives behave near the liquid-vapor critical point with temperature $T_\mrm{c}$ and pressure $P_\mrm{c}$. The theory of critical phenomena teaches us that many thermodynamic quantities follow asymptotically a power-law behaviour when approaching $T_\mrm{c}$ from above along the critical isochore $\rho=\rho_\mrm{c}$, with for instance:
\begin{equation}
\CV \sim A^+_\mrm{\CV} \tau^{-\alpha} , \quad \CP \sim A^+_\mrm{\CP} \tau^{-\gamma} , \quad \kT \sim A^+_\mrm{\kT} \tau^{-\gamma} , \quad \textrm{where} \quad \tau = \frac{T}{T_\mrm{c}} - 1 >0\; .
\label{eq:critexp}
\end{equation}
Renormalization group theory for the 3D Ising model gives $\alpha=0.1096$ and $\gamma=1.2373$~\cite{Campostrini_25thorder_2002}. It follows that, along the critical isochore:
\begin{eqnarray}
\kS & = & \frac{\CV}{\CP} \kT \sim \frac{A^+_{\CV} A^+_{\kT}}{A^+_{\CP}} \; \tau^{-\alpha} \rightarrow +\infty \quad\textrm{when}\quad \tau \rightarrow 0^+ \; ,\\
c & = & \frac{1}{\sqrt{\rho_\mrm{c} \kS}} \sim \sqrt{\frac{A^+_{\CP}}{\rho_\mrm{c} A^+_{\CV} A^+_{\kT}}} \; \tau^{\alpha/2} \rightarrow 0 \quad\textrm{when}\quad \tau \rightarrow 0^+ \; .
\end{eqnarray}
Consequently, when heating along the critical isobar at $P_\mrm{c}$ towards $T_\mrm{c}$, $\kS$ increases and $c$ decreases, which implies that there is a temperature interval just below $T_\mrm{c}$ where:
\begin{equation}
\left( \frac{\partial \kS}{\partial T} \right)_P (T,P_\mrm{c}) > 0 \quad\mrm{and}\quad \left( \frac{\partial c}{\partial T} \right)_P (T,P_\mrm{c}) < 0 \; .
\label{eq:dkSdT-dcdT2}
\end{equation}
From Eqs.~\ref{eq:dkSdT-dcdT} and \ref{eq:dkSdT-dcdT2}, it follows by continuity that there are lines in the $T-P$ plane, between the negatively-sloped TMD line and the liquid-vapor critical point, on which the temperature derivatives of $\kS$ and of $c$ along isobars vanish. This proves the existence of lines of extrema in $\kS$ and $c$ along isobars. It does not seem possible to conclude in general about whether the extrema are minima or maxima. However, in the particular case where the negatively sloped TMD line extends up to pressure above $P_\mrm{c}$ (which is the case for water~\cite{Holten_thermodynamics_2012}), the isobar at $P_\mrm{c}$ shows a minimum in $\kS$ and a maximum in $c$, proving the existence of a line of minima in $\kS$ along isobars and of a line of maximum in $c$ along isobars. These lines are indeed known in water at positive pressure.\\

\textbf{Comment on the order of the lines of extrema along isobars}

Here we discuss the relative positions of the lines of extrema in $c$, $\kT$, and density along isobars, for pressures where the TMD line has negative slope. From Mayer's relation, we have:
\begin{equation}
\frac{1}{c^2} = \rho \kS = \rho \frac{\CV}{\CP} \kT = \rho \left(1 - \frac{TV {\alpha_P}^2}{\kT} \right) \kT\; .
\end{equation}
The first two factors in the right hand side of the last equation reach a maximum along an isobar on the TMD, and so does their product $\Pi = \rho [1 - (TV {\alpha_P}^2 /\kT)] >0$. Therefore $(\partial \Pi/\partial T)_P >0$ for $T<T_\mrm{TMD}(P)$ and $<0$ for $T>T_\mrm{TMD}(P)$.
Taking the derivative:
\begin{equation}
\left( \frac{\partial (1/c^2)}{\partial T}\right)_P = \left( \frac{\partial \Pi}{\partial T}\right)_P \kT + \Pi \left( \frac{\partial \kT}{\partial T}\right)_P  \; .
\label{eq:dinvc2dT_P3}
\end{equation}
Along the negatively-sloped part of the TMD line, $(\partial (1/c^2) / \partial T )_P < 0$ (Eq.~\ref{eq:mastereq}) and $(\partial \kT / \partial T )_P <0$ (Eq.~\ref{eq:Sastry}). For $T>T_\mrm{TMD}$, $(\partial (1/c^2) / \partial T )_P < \Pi (\partial \kT / \partial T )_P$ so that the first derivative to reach zero upon heating from the TMD is $(\partial \kT / \partial T )_P$. Conversely, for $T<T_\mrm{TMD}$, $(\partial (1/c^2) / \partial T )_P > \Pi (\partial \kT / \partial T )_P$ so that the first derivative to reach zero upon cooling from the TMD is $(\partial (1/c^2) / \partial T )_P$. We conclude that, increasing temperature along an isobar, we meet the lines of extrema in the following order: LM$\kT|P$, Lm$c|P$ (if these two lines exist), TMD line, Lm$\kT|P$, and LM$c|P$.

A similar reasoning applied to the equations $1/c^2 = \rho \kS$ (see Eq.~\ref{eq:dinvc2dT_P}) and $\kS = [1 - (TV {\alpha_P}^2 /\kT)] \kT$ gives the following order: LM$\kT|P$, LM$\kS|P$, Lm$c|P$ (if these three lines exist), TMD line, Lm$\kT|P$, Lm$\kS|P$, and LM$c|P$.\\

\textbf{Lines of extrema implied by the existence of a TMD line with a maximum temperature}

Here we consider a fluid exhibiting a TMD line with a maximum temperature, i.e. a TMD line which has in the $T-P$ plane a negative slope at high pressure, and a positive slope at low pressure.

Using Eq.~\ref{eq:dPdT_V}, we write:
\begin{equation}
\left( \frac{\partial \kT}{\partial T} \right)_V = \left( \frac{\partial \kT}{\partial T} \right)_P + \left( \frac{\partial \kT}{\partial P} \right)_T \left(\frac{\partial P}{\partial T} \right)_V = \left( \frac{\partial \kT}{\partial T} \right)_P \; \mrm{at \; the \; TMD} \; .
\label{eq:dkTdT_V}
\end{equation}

Along the part of the TMD line with a negative slope in the $T-P$ plane, we know from Eqs.~\ref{eq:Sastry}, \ref{eq:mastereq}, and \ref{eq:dkTdT_V} and that:
\begin{equation}
\left( \frac{\partial \kT}{\partial T} \right)_P = \left( \frac{\partial \kT}{\partial T} \right)_V < 0 \; , \left( \frac{\partial \kS}{\partial T} \right)_P = \left( \frac{\partial \kS}{\partial T} \right)_V < 0 \; , \quad\mrm{and}\quad \left( \frac{\partial c}{\partial T} \right)_P = \left( \frac{\partial c}{\partial T} \right)_V > 0 \; .
\label{eq:negTMD}
\end{equation}

Along the part of the TMD line with a positive slope in the $T-P$ plane, these inequalities are reversed:
\begin{equation}
\left( \frac{\partial \kT}{\partial T} \right)_P = \left( \frac{\partial \kT}{\partial T} \right)_V > 0 \; , \left( \frac{\partial \kS}{\partial T} \right)_P = \left( \frac{\partial \kS}{\partial T} \right)_V > 0 \; , \quad\mrm{and}\quad \left( \frac{\partial c}{\partial T} \right)_P = \left( \frac{\partial c}{\partial T} \right)_V < 0 \; .
\label{eq:posTMD}
\end{equation}
By continuity, Eqs.~\ref{eq:negTMD} and \ref{eq:posTMD} imply the existence of lines of extrema in $\kT$, $\kS$, and $c$ along isobars, and of lines of extrema in $\kT$, $\kS$, and $c$ along isochores. Because the continuity argument is valid on any continuous path, these lines are found on both sides of the TMD: in the region limited above by the negatively-sloped part of the TMD line, and below by the positively-sloped part of the TMD line, and outside this region (towards high temperatures). All the derivatives in Eqs.~\ref{eq:negTMD} and \ref{eq:posTMD} vanish at the turning point of the TMD line, which therefore corresponds to an extremum of $\kT$, $\kS$ and $c$ along isobars and isochores. The type of extremum (minimum or maximum) is not fully determined without more assumptions. For instance, it was shown in Ref.~\cite{Sastry_singularityfree_1996} that three cases were possible for $\kT$: LM$\kT|P$ extending above the TMD turning point until it merges with the Lm$\kT|P$, Lm$\kT|P$ extending below the TMD turning point until it merges with the LM$\kT|P$, or Lm$\kT|P$ extending down to zero temperature. The additional existence of a liquid-liquid critical point implies additional properties as shown in the next section.\\

\textbf{Lines of extrema implied by the existence of a liquid-liquid critical point}

Here we consider a fluid exhibiting a liquid-liquid critical point at temperature $T_\mrm{c,2}$ and pressure $P_\mrm{c,2}$. We further assume that the corresponding liquid-liquid transition belongs to the Ising 3D universality class, and its slope is negative in the $T-P$ plane. Following the reasoning made near the liquid-vapor critical point in section \textit{Existence of lines of extrema in $\kS$ and $c$ along isobars above the TMD\ldots}, we find that $\kT$, $\CP$, and $\kS$ diverge at the liquid-liquid critical point, while $c$ reaches 0. We deduce the following properties. Approaching $T_\mrm{c,2}$ from below along the critical isobar at $P_\mrm{c,2}$, there is a temperature interval just below $T_\mrm{c,2}$ where:
\begin{equation}
\left( \frac{\partial \kT}{\partial T} \right)_P (T,P_\mrm{c,2}) > 0 \; , \left( \frac{\partial \kS}{\partial T} \right)_P (T,P_\mrm{c,2}) > 0 \; , \quad\mrm{and}\quad \left( \frac{\partial c}{\partial T} \right)_P (T,P_\mrm{c,2}) < 0 \; .
\label{eq:signPc2}
\end{equation}

Approaching $T_\mrm{c,2}$ from above along the critical isobar at $P_\mrm{c,2}$, there is a temperature interval just above $T_\mrm{c,2}$ where:
\begin{equation}
\left( \frac{\partial \kT}{\partial T} \right)_P (T,P_\mrm{c,2}) < 0 \; , \left( \frac{\partial \kS}{\partial T} \right)_P (T,P_\mrm{c,2}) < 0 \; , \quad\mrm{and}\quad \left( \frac{\partial c}{\partial T} \right)_P (T,P_\mrm{c,2}) > 0 \; .
\label{eq:signPc2above}
\end{equation}
When heating along a path going from $(T_\mrm{c,2}^-,P_\mrm{c,2})$ to $(T_\mrm{c,2}^+,P_\mrm{c,2})$ under the critical point, Eqs.~\ref{eq:signPc2} and \ref{eq:signPc2above} show that the above derivatives change sign. This implies by continuity the existence of a LM$\kT|P$, LM$\kS|P$, and Lm$c|P$, all converging towards the liquid-liquid critical point. A similar reasoning leads to the existence of a line of maxima in $\CP$ along isotherms (LM$\CP|T$) converging towards the liquid-liquid critical point. The LM$\kT|P$ and LM$\CP|T$ had already been identified as consequences of the liquid-liquid critical point~\cite{Poole_phase_1992,Poole_density_2005}. Similar to the fact that the line of extrema in $\kT$ along isobars intersects the TMD line at its maximum temperature~\cite{Sastry_singularityfree_1996}, it was also shown that the line of extrema in $\CP$ along isotherms intersects the TMD line at its minimum pressure~\cite{Poole_density_2005}.\\

\textbf{Lines of extrema implied by the existence of a TMD line with a maximum temperature and a liquid-liquid critical point}

Here we consider a fluid exhibiting a liquid-liquid critical point at temperature $T_\mrm{c,2}$ and pressure $P_\mrm{c,2}$ as in the previous section, and also having a 
TMD with a maximum temperature, i.e. a TMD line which has in the $T-P$ plane a negative slope at high pressure, and a positive slope at low pressure. We further assume that the liquid-liquid critical point is inside the region enclosed by the TMD.

Because of the negative slope of the liquid-liquid equilibrium line in the $T-P$ plane, the critical isochore also has a negative slope. We have seen above that $\kT$, $\CP$, and $\kS$ diverge at the liquid-liquid critical point, while $c$ reaches 0. Approaching $T_\mrm{c,2}$ from above along the critical isochore with volume $V_\mrm{c,2}$, there is a temperature interval just above $T_\mrm{c,2}$ where:
\begin{equation}
\left( \frac{\partial \kT}{\partial T} \right)_V (T,P(T,V_\mrm{c,2})) < 0 \; , \left( \frac{\partial \kS}{\partial T} \right)_V (T,P(T,V_\mrm{c,2})) < 0 \; , \quad\mrm{and}\quad \left( \frac{\partial c}{\partial T} \right)_V (T,P(T,V_\mrm{c,2})) > 0 \; .
\label{eq:signVc2}
\end{equation}

The TMD line has a part with a positive slope in the $T-P$ plane where Eq.~\ref{eq:posTMD} holds. Therefore Eqs.~\ref{eq:posTMD} and \ref{eq:signVc2} imply by continuity the existence of lines of extrema in $\kT$, $\kS$ and $c$ along isochores, in the region circled by (i.e. at temperature below) the TMD line. These lines of extrema intersect the TMD line at its turning point. The nature of the extrema (maxima or minima) is undetermined. We have of course that along an isochore, because $c=1/\sqrt{\rho \kS}$, a maximum in $\kS$ is a minimum in $c$ and vice-versa.

We note that the lines of extrema in $\kT$, $\kS$ and $c$ along isochores do not converge towards the liquid-liquid critical point, because the relevant derivatives are non-zero there (Eq.~\ref{eq:signVc2}). The position of the lines of extrema relative to the liquid-liquid critical point depend on the pressure of the TMD turning point. If the TMD turning point is at a pressure above $P_\mrm{c,2}$, Eq.~\ref{eq:posTMD} holds on the TMD at $P_\mrm{c,2}$. Combining Eqs.~\ref{eq:posTMD} and \ref{eq:signVc2} shows that the lines of extrema exist at $P_\mrm{c,2}$, passing through a point with temperature above $P_\mrm{c,2}$. If instead the TMD turning point is at a pressure below $P_\mrm{c,2}$, Eq.~\ref{eq:negTMD} holds on the TMD at $P_\mrm{c,2}$. Comparing Eqs.~\ref{eq:negTMD} and \ref{eq:signVc2} then shows that the lines of extrema cannot be above $T_\mrm{c,2}$ at $P_\mrm{c,2}$ (except for multiple changes of sign of the derivatives along the isobar at $P_\mrm{c,2}$). An example of the latter case is given by the TIP4P/2005 (Fig.~\ref{fig:TIP4P2005}), where LM$\kT|\rho$ and Lm$c|\rho$ are found: the two lines end on the liquid-liquid equilibrium line, passing to the left of the liquid-liquid critical point.


\end{document}